\documentclass[fleqn,10pt]{wlscirep}
\usepackage[utf8]{inputenc}
\usepackage[T1]{fontenc}
\usepackage{bm}
\usepackage{graphicx}
\title{Heat diffusion related damping process in highly precise coarse-grained model for SWCNT's nonlinearity}

\author[1,*]{Heeyuen Koh}
\author[2]{Chiashi Shohei}
\author[2]{Junichiro Shiomi}
\author[2,*]{Shigeo Maruyama}

\affil[1]{Mechanical and Aerospace Engineering Department, Seoul National University, 1 Gwanak-ro, Gwanak-gu, Seoul, 08826, South Korea}
\affil[2]{Mechanical Engineering Department, The University of Tokyo, Department of Mechanical Engineering, 7-3-1 Hongo, Bunkyo-ku, Tokyo 113-8656, Japan }

\affil[*]{hy\_koh@snu.ac.kr}
\affil[*]{maruyama@photon.t.u-tokyo.ac.jp}
\begin{abstract}
Second sound and heat diffusion in single-walled carbon nanotubes (SWCNT) is well-known phenomena which is related to the high thermal conductivity of this material. In this paper, we have shown that the heat diffusion along the tube axis affects the macroscopic motion of SWCNT and adapting this phenomena to coarse-grained model can improve the precision of the coarse-grained molecular dynamics (CGMD) exceptionally.  The nonlinear macroscopic motion of SWCNT in the free thermal vibration condition in adiabatic environment is demonstrated in the most simplified version of CG modeling as maintaining finite temperature and total energy with suggested dissipation process derived from internal heat diffusion. The internal heat diffusion related to the cross correlated momentum from different potential energy functions is considered, and it can reproduce the nonlinear dynamic nature of SWCNTs without external thermostatting in CG model. Memory effect and thermostat with random noise distribution are not included, and the effect of heat diffusion on memory effect is reviewed through Mori-Zwanzig formalism. This diffusion shows perfect syncronization of the motion between that of CGMD and MD simulation, which is started with initial conditions from the molecular dynamics (MD) simulation. The heat diffusion related to this process has shown the same dispersive characteristics to second wave in SWCNT. This replication with good precision indicates that the internal heat diffusion process is the essential cause of the nonlinearity of the tube. The nonlinear dynamic characteristics from the various scale of simple beads systems are examined with expanding its time step and node length simultaneously. 
\end{abstract}

\begin{document}

\flushbottom
\maketitle

\thispagestyle{empty}

\section*{Introduction}

 The thermal energy has a significant importance on the dynamics of the system in the scale of nm$\sim \mu$m. Understanding how random fluctuation \cite{Arroyo2004a,Zhigilei2005a,Strachan2005,Hijon2006,Shell2008} from thermal energy operates the dynamics in this range could expand the capability of the simulation model which should compromise the detailed dynamics from atomic scale phenomena. \cite{Strachan2005,Espanol2011,Espanol2016}. The effort for parameterization and to reveal the detailed mechanism on hierarchical structures\cite{Buehler2008,Elliott2011,Jebahi2016,Karatrantos2016,Jung2019,Li2019,Voth2017,Espanol2017} often reaches to continuum scale expressions as an effective descriptor.\cite{Li2016,Jung2017,Rudd1998,Rudd2005,Eom2011} The validation of these trials has been demonstrated its capability to delineate the role of thermal motion in macroscale through the comparison of phonon dispersion relations,\cite{Rudd1998,Rudd2005,Li2019} which shows the coarse-grained description can manage thermal condition in atomic scale. The theoretical approach to depict the memory effect and random noise distribution in more precise process for various simulation scales and models is on-going research. \cite{Ma2016,Chu2017}.

 Coarse-grained modeling for single-walled carbon nanotubes(SWCNT) has been an object to study thermal energy influence to macroscopic motion and morphology of complex made as a composites for theoretical and practical application. Quantified validation of coarse- grained (CG) modeling has tried in various methods, mostly in static characteristics. \cite{Zhigilei2005a,Jacobs2012,Buehler2006,Volkov2010,Volkov2012,Wittmaack2018,Wittmaack2019} The parameters for coarse graining of SWCNT has been fully explored with various size of SWCNT, \cite{Zhigilei2005a,Jacobs2012,Buehler2006} and the dynamic characteristics in acoustic dissipation from global deformation has been investigated.\cite{Jacobs2012} On the other hands, CGMD simulations suggesting the structure of SWCNT complex have been achieved for thermal conductivity and mechanical properties of the complex which are severely dependent on the characteristics of morphology.\cite{Volkov2010,Volkov2012,Won2013} For example, there is CGMD simulation of Won et al.,\cite{Won2013} which has made the CG modeling for vertically-aligned SWCNT forest (VA-SWCNT) by top-down method. The structure directly duplicated from SEM images has allowed to simulate the role of each structure type of VA-SWNT forest successfully. Further research on the multi-scale modeling with good efficiency and precision of VA-SWCNT forest\cite{Wittmaack2018,Wittmaack2019} proves that even the dynamic replication of VA-SWCNT in chemical vaporized decomposition (CVD) and further process on it\cite{Cui2013} are practicable. 

The CG modeling for saving the computational expenses of atomic simulation directly means that the trial should lose its detailed dynamic characteristics caused by such condition. Most progressed CGMD simulation for morphology, such as VA-SWCNT forest\cite{Wittmaack2018,Wittmaack2019} or buckypaper,\cite{Volkov2010,Volkov2012} has a cylinder shaped CNT to keep realistic dynamic and structural characteristics with taking computational expanse. Some exceptions are dissipative particle dynamics (DPD) modeling with polymers\cite{Liba2008,Maiti2008,Wang2011} and the mathematical random network of the sparse entanglement.\cite{Schiffres2012} In other studies as well,\cite{Li2019,Peter2009} composing coarse grained structure to maintain the dynamic features of individual molecule at certain level is essential to enhance the methodology to analyze in multiscale systems.\cite{Volkov2010,Muller2013,Kolomeisky2013} 

Recently, Koh et al.\cite{Koh2015} have reported that the bending motion of SWCNT in thermal equilibrium condition has nonlinear characteristics from MD simulations. The whirling motion of SWCNT repeatedly appears in the course of conventional planar bending motion. This whirling motion also changes its rotational direction in each appearance. Successful CG modeling of SWCNT would show the same motion characteristic: the nonlinear bending motion as reported.\cite{Koh2015,Liu2015} Any molecule which has one dimensional shape with fixed ends is suspected to have the similar nonlinear macroscopic motion characteristics according to the theoretical approach. The research scope of this paper is focused on a better algorithm for the simple beads model, which is the most simplified version of SWCNT. If the understanding on the role of thermal randomness to macroscopic motion is clear, it would reduce the complexity of dynamics into a simple beads system and it could be validated through CGMD simulation. The result of this trial would give the causality of nonlinear dynamic characteristics in SWCNT so that it could be manageable at some level of engineering.

\begin{figure}[ht]
\includegraphics[scale=0.5]{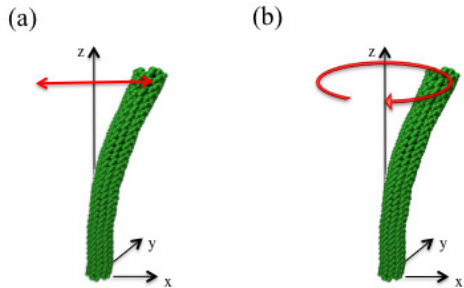}
\caption{Visualization of motion from MD simulation. The averaged coordinate of each carbon ring perpendicular to tube axis is amplified 5 times to clearly visualize the motion: (a) planar bending motion, (b) non planar motion.}\label{fig:Fig1}
\end{figure}

\section*{Results}
\subsection*{ Coarse-grained modeling for nonlinear motion in SWCNT}

The key feature of the nonlinear macroscopic motion in SWCNT is the motion type exchange.\cite{Ho1976,Koh2015,Liu2015} Each motion type as represented in Fig. \ref{fig:Fig1}, should be appeared repeatedly if the designed CG modeling has identical dynamic characteristics which has been shown in several atomic system in simulation\cite{Koh2015,Liu2015} and experiments.\cite{Barnard2019}  Judging by the high Q factor of SWCNT, quadratic functions for bond length and angle potential energy to model them as a harmonic potential energy seems reasonable, but it does not assure the nonlinearity like the motion type exchanges as shown in atomic scale system since it is defined in a plane.

  Governing equation for nonlinear bending\cite{Koh2015} describes the motion characteristics of SWCNT which is coarse-grained along tube axis-wise and its analytic solution provided the nonlinearity including motion exchange as shown in the MD simulation. For this nonlinear bending equation derived from Green Lagrangian strain definition, the same strain definition can be applied to coarse-grained model. According to the governing equation \cite{Koh2015}, the nonlinear behavior is the result from the combination of the bending on two perpendicular planes and length-wise deformation composed of a single quadratic function. Thus, the Hamiltonian for SWCNT which has identical to the Green-Lagrangian strain can be expressed as below:

\begin{eqnarray}
H_{MD}(x_1,...,x_n,p_1,...,p_n) =   \frac{1}{2} \frac{p^T p}{m} + \Phi_{MD}(x_1,...,x_n),\label{eqn:Eq1}  \\
\frac{dx_i}{dt} = \frac{\partial H}{\partial p_i}, \label{eqn:Eq2} \\
\frac{dp_i}{dt} = -\frac{\partial H}{\partial x_i}, \label{eqn:Eq3} \\
\Phi_{MD} = \Phi_{G.E.} + \Phi_{int},\label{eqn:Eq4} \\
\Phi_{G.E.} = \frac{1}{2}k \left( w_l + w_{ \theta}\right)^2, \label{eqn:Eq5} 
\end{eqnarray}

,here, $x_i$ and $p_i$ are the displacement and velocity of the coarse grained particle, respectively. $\Phi_{G.E.}$ is the potential energy which is given by the Green-Lagrangian strain definition and $\Phi_{int}$ is the internal energy of coarse-grained particle which is ignored in governing equation. $\Phi_{G.E.}$ is defined as the quadratic function of the combination of bending i.e. angle deformation and its deformation along bond length-wise. Its full description is in Supporting material A.

Ideally, coarse-grained model should be composed of two independent Hamiltonian system for Stackel condition \cite{Pars1949} which indicates the momentum has to be separated into two independent variables according to each potential energy type which is defined for each type of deformation. 

\begin{eqnarray}
H_{CG} = H_{L} + H_{\theta},\label{eqn:Eq6}\\
H_{\l}(l_1,...,l_n,p_{L1},...,p_{Ln}) = \frac{1}{2} \frac{p_{L}^{T} p_{L}}{m} +  \Phi_{l}, \label{eqn:Eq7} \\
H_{\theta}(\theta_1,...,\theta_n,p_{\theta,1},...,p_{\theta,n}) =\frac{1}{2} \frac{p_{\theta}^{T} p_{\theta}}{m} + \Phi_{\theta} ,\label{eqn:Eq8} 
\end{eqnarray}

here, $H_L$ $H_{\theta}$ are Hamiltonian for bond-length and angle deformation, respectively. $p_L$ and $p_{\theta}$ are the momenta for each type of deformation. 

However, in case of Coarse-grained molecular dynamics (CGMD) using simple beads system, there is no momentum separation for each type of the potential energy which is defined in a separated manner:
 
 \begin{eqnarray}
H_{CG}(x_1,...,x_n,p_1,...,p_n) =   \frac{1}{2} \frac{p^T p}{m} + \Phi_{CG}(x_1,...,x_n), \label{eqn:Eq9} \\
\frac{dx_i}{dt} = \frac{\partial H}{\partial p_i}, \label{eqn:Eq10} \\
\frac{dp_i}{dt} = -\frac{\partial H}{\partial x_i},\label{eqn:Eq11}  \\
\Phi_{CG} = \Phi_{L} + \Phi_{\theta},\label{eqn:Eq12} 
\end{eqnarray}

The change from Eq. (\ref{eqn:Eq1})$\sim$ Eq.(\ref{eqn:Eq5}) to Eq. (\ref{eqn:Eq9})$\sim$Eq.(\ref{eqn:Eq12}) is in the definition of potential energy, whose form is two independent harmonic functions in CG modeling unlike that of MD and nonlinear governing equation. The difference between ideal CG model in Eq. (\ref{eqn:Eq6})$\sim$ Eq.(\ref{eqn:Eq8}) and Eq. (\ref{eqn:Eq9})$\sim$Eq.(\ref{eqn:Eq12}) is, however, that the momentum in both Hamiltonian system is integrated into a value.

\subsection*{ Influence of unseparated momenta in MZ formalism}

 It is very important to define the direction of $p_i$ and $\dot{\bm{q}}$, $ \bm{e}_{\dot{q}_i}$, precisely. Due to the time integration and sharing the momenta, the direction of each component of Eq. (\ref{eqn:Eq15}) is happened to be defined through the bond length $\ell$ and angle $\theta$ as defined in Eq. (\ref{eqn:Eq13}) which are the variable of the potential functions at each atom. It is easy to get the unit vectors and make the momentum $p_i$ can be divided into $p_{i,\ell}$ and  $p_{i,\theta}$ in orthogonal direction at a certain moment so that they are the functions of the bond length $\ell$ and angle $\theta$ at every moment. 
 
\begin{eqnarray}
\bm{p}_{i}=\left(p'^{\theta}_i+p'^{\ell}_i\right) \bm{e}_{p_i}\\ \label{eqn:Eq13}
        = p^{\theta}_i \bm{e}_{p_{\theta}}+p^{\ell}_i \bm{e}_{p_{\ell}}, \label{eqn:Eq13_} 
\end{eqnarray}

here, $ \bm{p_{i}}$ is the velocity of $i$ th unit mass in CGMD. $p_i^{\ell}$ and  $p_i^{\theta}$ are the momenta along the angle and bond length in CGMD simulation. They are the scalar components of total velocity so that they share the unit vector, $\bm{e}_{p_i }$ which is the direction of total velocity, $\bm{p}_i$. The components of $\bm{p}_i$ from $\bm{p}_i^{\ell}$ and $\bm{p}_i^{\theta}$ are noted as $p_i^{'\ell}$ and $p_i^{'\theta}$.
%In this way, the amount of cross correlated momentum could be shown.

 In case of CG simulation with two independent yet sharing the momenta, as written in Eq. (\ref{eqn:Eq13}). The Hamiltonian extended from modified principle of Hamilton becomes, 
 
 \begin{eqnarray} 
 \delta I = \delta \int_{t_1}^{t_2}(\dot{q}_i p_i -   H (q,\dot{q},t)),\label{eqn:Eq14} 
 \end{eqnarray}
 with, 
 \begin{eqnarray} 
 %\dot{\bm{q}_i} = (\dot{q}^{\theta}_{i} + \dot{q}^{\ell}_{i}) \bm{e}_{\dot{q}_i}, \label{eqn:Eq15} \\ 
  H (q,\dot{q},t) =   H_{\theta} +   H_{\ell}. \label{eqn:Eq16} 
 \end{eqnarray}
 
 When we suppose that $H_{\theta}$ and $H_{\ell}$ are two independent Hamiltonians, the least of action principle should be valid for each with the $\dot{\bm{q}}$. It is the change of displacement that a mass is experiencing for both of phase space of $H_{\theta}$ and $H_{\ell}$. It could be separated accordingly as shown in Eq. (\ref{eqn:Eq13}). %This is why $ \dot{q}^{\theta}_{i}$ and $ \dot{q}^{\ell}_{i}$ are noted as a scalar variable along the direction of total displacement change $\dot{\bm{q}}$ in Eq. (\ref{eqn:Eq15}). 

 With the Eq. (\ref{eqn:Eq14}) $\sim $Eq. (\ref{eqn:Eq16}), one could get a twist in Liouville theorem, especially on the infinitesimal volume in phase space for each type of Hamiltonian. For example, the volume of $  H_{\ell}$ at a certain moment $t$ is; 
 
  \begin{eqnarray}
  \dot{q} _i^{\ell}=\frac{\partial   H}{\partial p_i^{\ell}},\label{eqn:Eq17} \\
  \dot{p}_i^{\ell} =-\frac{\partial   H}{\partial q_i^{\ell}} - \gamma p_i^{\ell} - \gamma' \dot{q}_{i}^{\theta}, \label{eqn:Eq18} \\
  dq_i^{'\ell} dp_i^{'\ell}  =   dq_i^{\ell} dp_i^{\ell} \left[1+(\gamma p_i^{\ell} + \gamma' \dot{q}_{i}^{\theta}) \delta t \right], \label{eqn:Eq19}  
  \end{eqnarray}
 
 where $\gamma$ and $\gamma'$ is the proportionality of $\dot{q}_{\theta}$ and $p_i^{\ell}$ to $q_i^{\ell}$ in $  H_{\ell}$ from Taylor expansion. Note that the damping in Eq. (\ref{eqn:Eq19}) has two different term proportional to different type of momentum. Naturally given damping by adapting two independent Hamiltonian is not adjusted to specific temperature and given initial condition so that CGMD simulation is resulted at around 1000 K as shown in Supporting material A. It could be supposed as time varying variable, but we leave this value as constant, for now. Eq. (\ref{eqn:Eq19}) makes the volume in phase space non-conservative with additional damping on $p_i$ and $q_i$. To measure the consequence of the term $\gamma$, the projection operator method used in Mori-Zwanzig formalism is adapted from Kinjo and Hyodo\cite{Kinjo2007} and Kauzlaric\cite{Kauzlaric2011}.
 
  In atomic system which has microscopic state $z$ in the phase space $\hat{\Gamma}(t) = \{\hat{r}_{\alpha i},\hat{p}_{\alpha i}\}$, the Liouville operator, $L$ shows evolution through time\cite{Kauzlaric2011}: 

\begin{eqnarray}
\dot{z}(t) = L z,  \label{eqn:Eq20}  \\
z=exp(Lt) z_0. \label{eqn:Eq21} 
\end{eqnarray} 

The coarse-grained model that adapts CoM (Center of Mass) variables from $z$ would share the notation as followings: 

\begin{eqnarray}
%H=K+U \\
%K = \Sigma_{}^{} \Sigma_{}^{}\frac{}{}
\hat{R}_{\alpha} \equiv \frac{\Sigma_{i} m_{\alpha i}\hat{r}_{\alpha i}}{M_{\alpha}}, \label{eqn:Eq22}  \\
\hat{P}_{\alpha} \equiv \Sigma_i \hat{p}_{\alpha i},  \label{eqn:Eq23} \\
M_{\alpha}  \equiv \Sigma_i m_{\alpha i}, \label{eqn:Eq24} 
\end{eqnarray}

 here, $\alpha$ can be either $\ell$ or $\theta$ in the simple beads system. In Mori-Zwanzig formalism,\cite{Kauzlaric2011,Hijon2009,Kinjo2007} the evolution of variable for coarse-grained particle which is the function of a small group of microscopic state, $A_{\mu}=A_{\mu}(z(t))$ is managed with the phase space density $f_s(\hat{\Gamma}_s(t),\hat{\Gamma}_s)$ for Hamiltonian H and this can be expanded to the variables defined in CoM\cite{Kinjo2007}.

%, where each term in Eq. () is as below:  

%\begin{eqnarray} 
%e^{Lt}PLA = e^{Lt} \frac{(LA,A)}{(A,A)} A = \frac{(LA,A)}{(A,A)} e^{Lt}A =\Omega A(t),\\
%PLe^{QLt}QLA = PLF(t) = PLQF(t) = - \frac{ ( F(t),F(0))}{(A,A)} A = -K(t)A,
%\end{eqnarray}

Dynamic variable $g(\hat{\Gamma}(t))$ can be defined on the $f_s$ with the equilibrium distribution $\Psi(\hat{\Gamma})=e^{-\beta H}/Z$. The projection P and Q = 1-P can divide $g(\hat{\Gamma}(t))$ and $\left( \frac{d}{ds} \right)_{\Gamma} f_s$ as:
\begin{eqnarray} 
g(\hat{\Gamma}(t)) = g_P(\hat{\Gamma}(t)) + g_Q(\hat{\Gamma}(t)), \label{eqn:Eq25} \\
\left( \frac{d}{ds} \right)_{\Gamma} f_s(\hat{\Gamma_s}(t);\Gamma_s)  =P iLf_s \left(\hat{\Gamma}_s(t);\Gamma_s\right) +  Q iLf_s \left( \hat{\Gamma}_s(t);\Gamma_s \right),  \label{eqn:Eq26} 
\end{eqnarray} 

  where,
 \begin{eqnarray} 
 \left<f_s(\hat{\Gamma}_s(t_0);\Gamma_s) g_Q(\hat{\Gamma}(t)) \right> =0, \label{eqn:Eq27} \\
 \left(A,B \right) \equiv  \left<A(\hat{\Gamma)},B(\hat{\Gamma}) \right> =  \int d\hat{\Gamma} A(\hat{\Gamma})B(\hat{\Gamma})\Psi(\hat{\Gamma}).   \label{eqn:Eq28} 
 \end{eqnarray} 
  
 The integration of Eq. (\ref{eqn:Eq27}) is on Hamiltonian $H$, and the projector $P$ of $  H_{\theta}$, which is the one of component of Hamiltonian $H$, is on $f_{s\ell}$ which is; 
 
 \begin{eqnarray}
  f_{s\ell}(\hat{\Gamma}(t);\Gamma_s) \equiv \delta (\hat{\Gamma}_{s\ell}(t)-\Gamma_s) = \prod_{\alpha} \delta (\hat{R}_{\alpha}-R_{\alpha}) \delta(\hat{P}_{\alpha}-P_{\alpha}).  \label{eqn:Eq29} 
 \end{eqnarray}   
  
  , where $\alpha$ is for the CoM variable for $\ell$ and $\alpha'$ for $\theta$. The time evolution of $ f_{s\ell}$ is defined with its Hamiltonian; 
  
  \begin{eqnarray}
\left(\frac{d}{dt} \right)_{\Gamma} f_s = -\sum_{\alpha} \sum_{i} \bigg\{ \frac{\partial H}{\partial \hat{r}_{\alpha i}} \cdot \frac{\partial}{\partial \hat{p}_{\alpha i}} -  \frac{\partial H}{\partial \hat{p}_{\alpha i}} \cdot \frac{\partial}{\partial \hat{r}_{\alpha i}} \bigg\} f_s \nonumber  \\
=  - \sum_{\alpha}  \bigg\{ \hat{F'}_{\alpha} \cdot \frac{\partial}{\partial \hat{P}_{\alpha}} -  \frac{\hat{P'}
_{\alpha}}{M_{\alpha}} \cdot \frac{\partial}{\partial \hat{R}_{\alpha}} \bigg\} f_s \equiv iL_s f_s, \label{eqn:Eq30} 
\end{eqnarray}

where,

\begin{eqnarray}
L_s \equiv -\sum_{\alpha} \bigg\{ \hat{F'}_{\alpha} \cdot \frac{\partial}{\partial \hat{P}_{\alpha}} -  \frac{\hat{P'}
_{\alpha}}{M_{\alpha}} \cdot \frac{\partial}{\partial \hat{R}_{\alpha}}  \bigg\}, \label{eqn:Eq31} \\
\hat{F'}_{\alpha} = -\sum_{n_{\alpha}}^{i} \frac{\partial U}{\partial \hat{r}_{\alpha i}} +\gamma \hat{P}_{\alpha}+\gamma' \dot{q}_{\theta},\label{eqn:Eq32} \\
\hat{P'}_{\alpha} =\hat{P}_{\alpha}.  \label{eqn:Eq33} 
\end{eqnarray}

%The real system which multiple Hamiltonian would be constructed with Bogoliubov-Born-Green-Kirkwood-Yvon hierarchy.   
To clarify the influence of $\gamma \dot{q}_{\theta}\hat{P}_{\alpha}$ and $\gamma' \dot{q}_{\theta}$, lets suppose we have $H_{\ell}$ and $  H_{\theta}$ only with rigid N CG particles as defined by Eq. (\ref{eqn:Eq9})$\sim$ Eq.(\ref{eqn:Eq13})  and being untacted by other influence like thermal energy. In this way, $i   L_s$ in Eq. (\ref{eqn:Eq30}) can be divided with operator $ P$ has additional damping $\gamma \hat{P}_{\alpha}\frac{\partial}{\partial \hat{P}}$ from which is as defined in ref.\cite{Kinjo2007} and $Q$ which has very specific definition, $\gamma \dot{q}_{\theta}  \frac{\partial}{\partial \hat{P}}$. The term for fluctuation force and memory effect are changed so that the equation of motion from time evolution equation for phase-space density of $f_{s\ell}$ becomes;
  
  \begin{eqnarray} 
 %\left(\frac{d}{dt} \right)_{\Gamma} f_{s\ell} (\hat{\Gamma}_s(t);\Gamma_s)=\int d \Gamma'_s f_{s\ell}() i \Sigma() + \int^{t}_{0} d\tau \int x + ,
 \frac{d}{dt}\hat{  P}_{\sigma}=\frac{1}{\beta}\frac{\partial}{\partial \hat{R}_{\sigma}} \ln \omega(\hat{  R}) - \gamma \hat{P}_{\alpha}- \beta \sum_{\alpha}\int^{t}_{0} ds<\left[\delta P^{   Q}_{\sigma}(t-s)\right] \times \left[\delta P^{   Q}_{\alpha}(0) \right]^{T}  > \frac{\hat{P}_{\alpha}(s)}{M_{\alpha}} + \delta P^{Q}_{\sigma}(t), \label{eqn:Eq34} 
  \end{eqnarray} 
    
    where, 
   
  \begin{eqnarray}     
    P(\hat{\Gamma}(0),\Gamma_s) = -\sum_{\alpha} \big[ \dot{q}_{i,\theta}\cdot \frac{\partial}{\partial   P_{\alpha}} \delta(\hat{\Gamma}_s-\Gamma_s)\big]  \label{eqn:Eq35} \\ 
\int d \Gamma_{s}   P_{\sigma}   P(\hat{\Gamma}(0),\Gamma_s) = \delta P_{\sigma}=\gamma \dot{q}_{\theta}, \label{eqn:Eq36} \\
 \delta P^{Q}_{\sigma}(t) = e^{QiLt}  F(\hat{\Gamma}(0),\Gamma_s).  \label{eqn:Eq37} 
   \end{eqnarray}

 More specific derivation is in Method section. Note that $\delta P^{Q}_{\sigma}(t)$ is $\delta F^{Q}_{\sigma}(t)$ in the reference\cite{Kinjo2007}. It is changed to reveal that the fluctuation is from the momentum of other Hamiltonian.

  The second term on Eq. (\ref{eqn:Eq36}) becomes more specifically as below:  
 \begin{eqnarray} 
 - \beta \sum_{\alpha} \int^{t}_{0} ds \left<\left[ \delta P^{   Q}_{\sigma}(t-s)\right] \times \left[\delta P^{   Q}_{\alpha}(0) \right]^{T} \right> \cdot \frac{\hat{P}_{\alpha}(s)}{M_{\alpha}} = \nonumber  \\
 - \beta \sum_{\alpha} \left<  \left[ \int^{t}_{0} ds \delta P^{   Q}_{\sigma}(t-s) \cdot \frac{\hat{P}_{\alpha}(s)}{M_{\alpha}} \right] \times \left[\delta P^{  Q}_{\alpha}(0) \right]^{T}  \right>, \label{eqn:Eq38} 
 \end{eqnarray}    
% where,
%\begin{eqnarray} 
%\delta P^{  \mathscr Q}_{\sigma}(\tau) \equiv e^{-  \mathscr QiL\tau}\delta P_{\sigma} ,\\
%\delta P \equiv \hat{P}_{\sigma} - \int \frac{1}{\beta} \frac{\partial}{\partial \hat{R}_{\sigma}} \ln\omega(\hat{R})). 
%\end{eqnarray}  

% The last term of Eq. (32) is cross correlation with momentum from fluctuation and the force from potential energy $U$ as given in Eq. (26). Unless the value of this term is ignorable, it would contradict to the assumption that the system has two independent Hamiltonian system satisfying Stackel condition. Thus, the memory effect in Eq. () is equilvalent to the first term in Eq. (33) and this can be noted as $\frac{d}{dt} M_{P^{  \mathscr Q}_{\sigma},\hat{P}_{\alpha}}(s,t)$ as: 

%\begin{eqnarray} 
%  \sum _{\alpha}  \frac{d}{dt} M_{P^{  \mathscr Q}_{\sigma},\hat{P}_{\alpha}}(s,t)=\\
%   \sum _{\alpha}  \int ds \frac{d}{ds} \left( - \beta \left<\left[ \int^{t}_{0} ds  \delta F^{ \mathscr Q}_{\sigma}(t-s)\right] \times \left[\delta F^{  \mathscr Q}_{\alpha}(0) \right]^{T} \right> \cdot \frac{\hat{P}_{\alpha}(s)}{M_{\alpha}}  \right) = \\
% \sum _{\alpha} \frac{d}{dt}   \int ds \left( - \beta K(t) \cdot \frac{\hat{P}_{\alpha}(s)}{M_{\alpha}}  \right),  \\
% K(t) = \left<\left[ \delta P^{  \mathscr Q}_{\sigma}(t)\right] \times \left[\delta F^{  \mathscr Q}_{\alpha}(0) \right]^{T} \right>. 
% \end{eqnarray} 
 The value of time integration in Eq.(\ref{eqn:Eq38}) has same value with to cross correlation of those variables in frequency domain: 
 
\begin{eqnarray} 
 M = \int^{t}_{0} ds \delta P^{  Q}_{\sigma}(t-s) \cdot \frac{\hat{P}_{\alpha}(s)}{M_{\alpha}}, \label{eqn:Eq39} \\
 M_{\ell}( \omega ) = \int dt e^{i\omega t} \int^{t}_{0} ds \delta P^{   Q}_{\sigma}(t-s) \cdot \frac{\hat{P}_{\alpha}(s)}{M_{\alpha}} = \delta P^{   Q}_{\sigma}( \omega) \hat{P}_{\alpha}( \omega ). \label{eqn:Eq40} 
% K_F(\omega)= \int dt' e^{i\omega t'}\left<\left[ \delta P^{  \mathscr Q}_{\sigma}(t)\right] \times \left[\delta F^{  \mathscr Q}_{\alpha}(0) \right]^{T} \right>. 
 % = \int d^N p d^N q p^{eq} F(0) F(\omega) A(\omega).   
 \end{eqnarray}

\begin{figure}%[\sidecaptionrelwidth][t]
\centering
\includegraphics[scale=0.5]{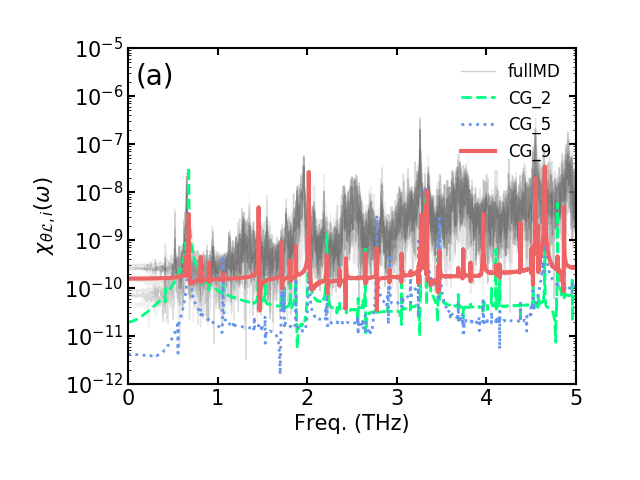}
\includegraphics[scale=0.5]{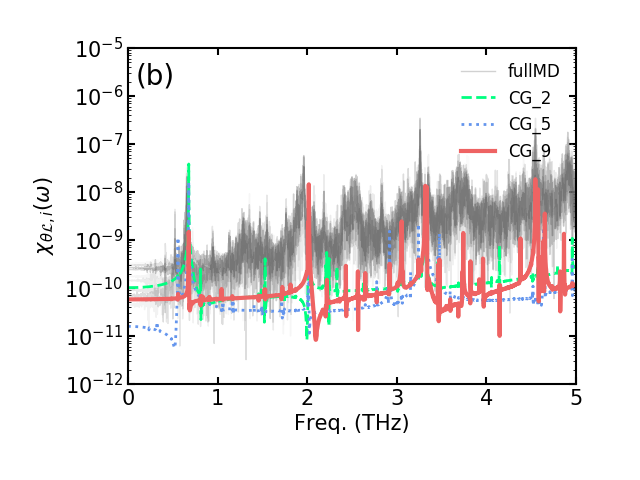}
\caption{Cross correlation of  from CGMD and simple beads model from MD data in frequency domain. Grey line is MD simulation duplicated with all CG particle in simple beads model, green, blue and red line are CGMD results at 2, 5, 9 th node: (a) without damping. MD simulation results has rather narrow range of distribution between each CG particle compared to CGMD results. The peak of CGMD results is close to delta function, (b) with internal heat diffusion. The difference between each node in CGMD is narrowed and the peak at certain frequency is diminished compared to (a). The shape of peaks is widened.}\label{fig:Fig2}
\end{figure}

$ M_{\theta}$ would be share the same definition and the value, which is managed in Fig. \ref{fig:Fig2}. Cross correlation in frequency domain is processed from CGMD and MD simulation data. The simulation conditions for MD and CGMD for (5,5) 8nm length SWCNT are in Method section. This quantity of CGMD is very distinguishably different to that of MD simulation whose results is averaged as simple beads system. No significant results have been found in the range below THz. The CGMD simulation has more distinctive peaks than that of MD simulation results. This means 1) In case of CGMD, it is almost identical condition to the rigid N CG particles. In case of MD simulation, the simple beads system which as displacement and velocity averaged from MD, the influence of thermal fluctuation or heat bath should be incorporated as another dissipation term, and 2) MD simulation has cross correlation which is not active in the bending frequency range but in optical mode range. For the integrity of the paper, auto-correlation of the momenta in each Hamiltonian is in Supporting material B showing non-Markovian characteristics. 

%In MD simulation, the force caused by $2\bm{\ell_i} \bm{\theta_i}$ is close to small perturbation in CG description level, and it could be disappeared as internal heat diffusion process as fluctuation-dissipation in THz range. 
% Q for orthogonal PE only and quantitative trial.. and then the results, heat bath is working, so that... changed in local/THz level interaction  To let the system have only two types of Hamiltonian which satisfies Stackel condition so that each system is ideally operating without any significant disturbances simultaneously, as the way they satisfy Stackel condition. 

\subsection*{Heat diffusion in General Langevin Equation}

 %From the Influence of heat bath 
It could be one interesting question how the thermal random motion from heat bath can help to secure the high Q factor of macroscopic motion as establishing two independent potential energy in their balance with the righteous level of cross correlated state. All these entangled situations have well condensed into a definition named free energy. The amount of the total force exerted on the macroscopic motion including its damping is regarded as the result of it.\cite{Crooks1999,Jarzynski1997} It is, however, not easy to define the value at each time step while the simulation is on running. 

\begin{figure}%[\sidecaptionrelwidth][t]
 \includegraphics[scale=0.3]{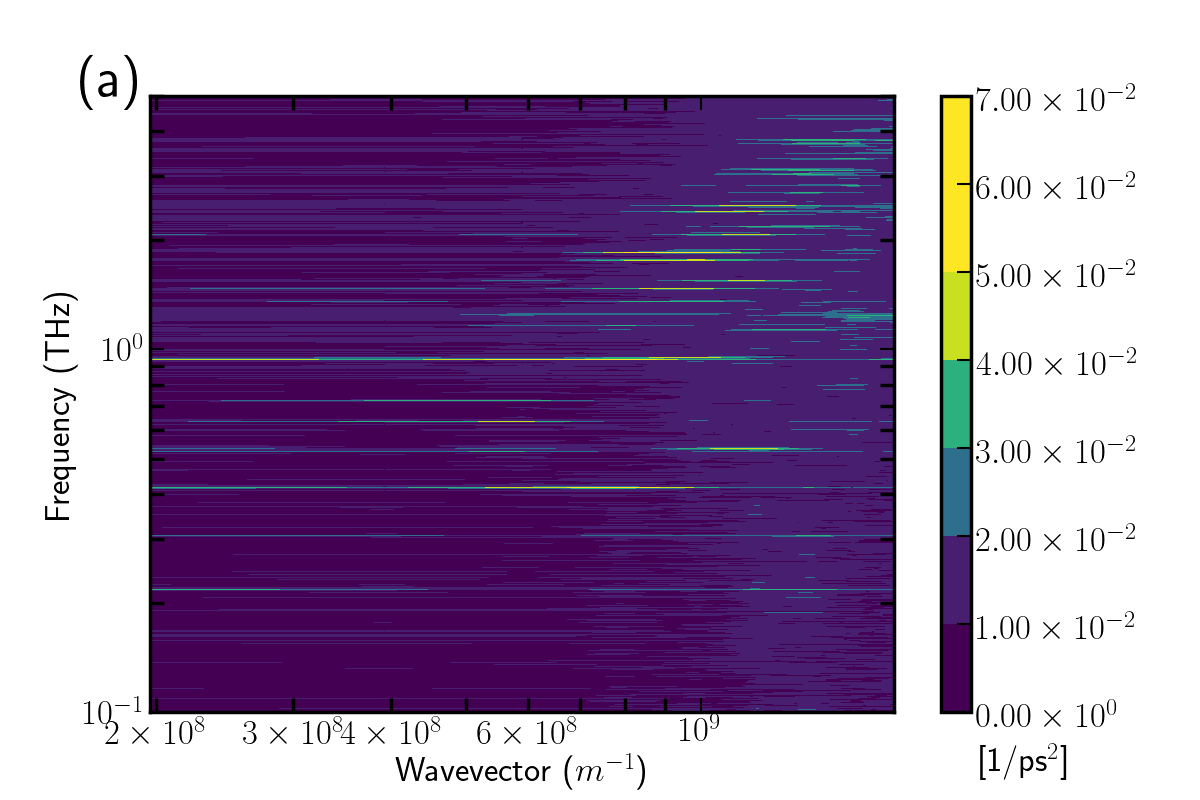}
 \includegraphics[scale=0.3]{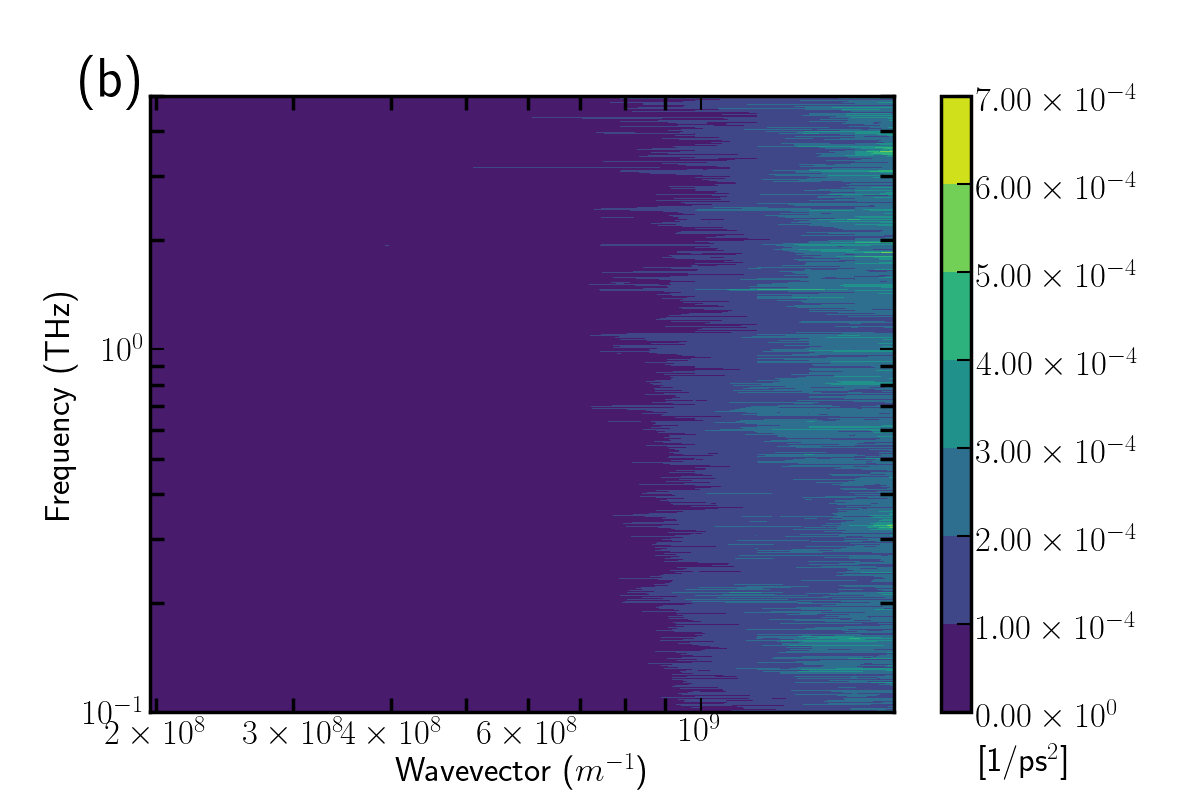}
 \includegraphics[scale=0.5]{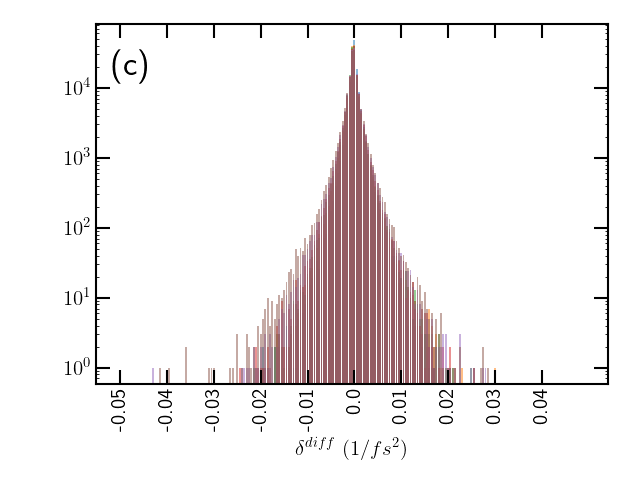}
 \includegraphics[scale=0.5]{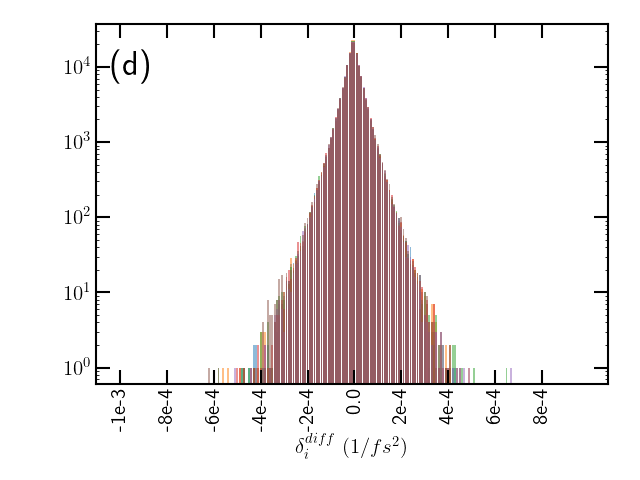} 
 
 \caption{The contour of the value of $\delta^{diff}$ along the tube axis. Each value is distributed with dt=10 fs along x axis. (a) MD simulation result, (b) CGMD simulation, (c) histogram of the diffused level in simple beads system from MD simulation result during 10 ns, the results from all UA are shown, (d) histogram for CGMD simulation results.} \label{fig:Fig3}
 \end{figure}

Effect of heat bath suggested by Zwanzig in 1961\cite{Zwanzig1961}, noted by additional Hamiltonian as $H_b$, is for the general influence of the collective dynamics from the Hamiltonian system belong to  the individual atom inside the CG particle. Since resolving the cross correlated state through heat bath will maintain the momentum and energy balance, this collective dynamics in atomic simulation can be regarded as a part of second wave modulation,$\partial^2 T / \partial t^2 = \partial^2 T / \partial z^2$, where $z$ is the variable for tube length axis. To adjust this process in CG model, it is presumed that small amount of the disturbance in kinetic energy of CG particle like  $K_{tot}=KE_0+\delta KE$ as a drift from stochastic motion of atoms in a CG particle affected by cross correlated state. From the direct proportionality between $KE$ and $T$, the amount of cross correlated state in momentum which is equivalent to $2p_l p_{\theta}$ from Eq. (\ref{eqn:Eq13}) can be regarded as the additional kinetic energy $\delta KE$. If this amount of energy is diffused as second wave, it should be shown through the dispersive characteristics of $\frac{ \partial ^2 p_{i,l} p_{i,\theta}}{\partial x ^2}$ in frequency range, by the value of modulation which can be projected into the slow variable in simple beads system and would be remained from averaging the displacement and velocity information from MD simulation. The actual proof to show the liason between presumed diffusion process and second wave is well shown in the dispersion plot in Fig.\ref{fig:Fig3} (a).It shows dispersive characteristics as explained by Lee and Lindsay 2017 \cite{Lee2017}. 

 The diffusion value, the results of second wave presumed to be related to diffusion process, $\frac{ \partial ^2 p_{i,l} p_{i,\theta}}{\partial x ^2}$. the perturbation proportional to this cross correlated state as a heat diffusion by the inter-molecular kinetic energy exchange will induce the desired consequence. 
 The system with this diffusion process could be written as below: 
 
\begin{eqnarray}
H_{CG} = H_{s}(X)+H_{b}(X,Y), \label{eqn:Eq41} \\
H(x_1,...,x_n,p_1,...,p_n) =   \frac{1}{2} \Sigma \frac{p_i^2}{m} + \Phi_{CG}(x_1,...,x_n), \label{eqn:Eq42}  \\
H_b (x_1,...,x_n,p_1,...,p_n) = \delta KE = \Sigma_i \left< \delta_i^{diff} \right>_{op}, \label{eqn:Eq43} \\
\delta_i^{diff} =\frac{ \partial ^2 p_{i,l} p_{i,\theta}}{\partial x ^2}.  \label{eqn:Eq44} 
\end{eqnarray} 

 This cross correlation can be incorporated as meta-dynamics derivation\cite{Gervasio2016} into the governing equation as defined in Eq. (\ref{eqn:Eq41}) $\sim$ Eq. (\ref{eqn:Eq44}) which is activated locally as a constrained term of Lagrangian multiplier. Further approximation and derivation are in the Supporting Information C. The Hamiltonian from this rough approximation is:

\begin{eqnarray}
m\ddot{l}+\alpha \left< \frac{\partial ^2 v_{\theta} }{\partial x^2} \right>_{op} = - \frac{\partial \Phi_{CG}}{\partial l},  \label{eqn:Eq45} \\
I\ddot{\theta}+\alpha' \left< \frac{\partial ^2 v_{\l} }{\partial x^2} \right>_{op} = - \frac{\partial \Phi_{CG}}{\partial \theta}, \label{eqn:Eq46}  
\end{eqnarray} 

here, To give this virtual force in THz range, the $+/-$ sign alternation is included which is noted as $ \left< \right>_{op}$  in  Eq. (\ref{eqn:Eq45}) $\sim$ Eq. (\ref{eqn:Eq46}). It is worthy to note that the time evolution of the momentum from projection operator method with heat diffusion process, which is embedded as a local heat diffusion $\partial^2 \dot{q}_{\theta or \ell}/ \partial x^2 \delta(\hat{\Gamma}_{i}-\Gamma_s)$ in optical mode $e^{i \omega_{op} t}$ involved instead of solid definition of momentum in Eq. (\ref{eqn:Eq45}) $\sim$ Eq. (\ref{eqn:Eq46}) makes the memory effect term $M$ in optical mode: 

\begin{eqnarray}
 M = \sum_{\alpha}\int^{t}_{0} ds \delta P^{  Q}_{\sigma}(t-s) \cdot \frac{\hat{P}_{\alpha}(s)}{M_{\alpha}}\\
 =  \int^{t}_{0} ds \dot{q}_{i}(t-s)e^{i \omega_{op} (t-s)} \cdot \frac{\hat{P}_{\alpha}(s)}{M_{\alpha}}. \label{eqn:Eq47} 
\end{eqnarray} 

 The time integration on optical mode could be a damping on the fluctuation $\delta P^{  Q}_{\sigma}(t)$, and the infinitesimal volume in phase space as written in Eq. (\ref{eqn:Eq19}) could be preserved with small oscillating. This virtual force is added to the conservative forces calculated from each potential energy so that it is included in both steps in velocity verlet algorithm.  

In Fig. \ref{fig:Fig3} (b), the same dispersion plot of $\frac{ \partial ^2 p_{i,l} p_{i,\theta}}{\partial x ^2}$ from CGMD calculated using Eq.(\ref{eqn:Eq45}) $\sim$ Eq.(\ref{eqn:Eq46}) has no dispersion at alll, which is very natural since the Equation of motion has artificially given perturbation in THz with the shorted wave length mode. The histograms of the value of diffusion in CGMD, however, have almost identical distribution with that in MD simulation, as shown in Fig. \ref{fig:Fig3} (c) and (d). This result would reflect that the similarity of dynamics in CGMD in the phase space with similar $p^{eq}$.

\subsection*{Validation}

\begin{figure}%[\sidecaptionrelwidth][t]
\centering
\includegraphics[scale=0.45]{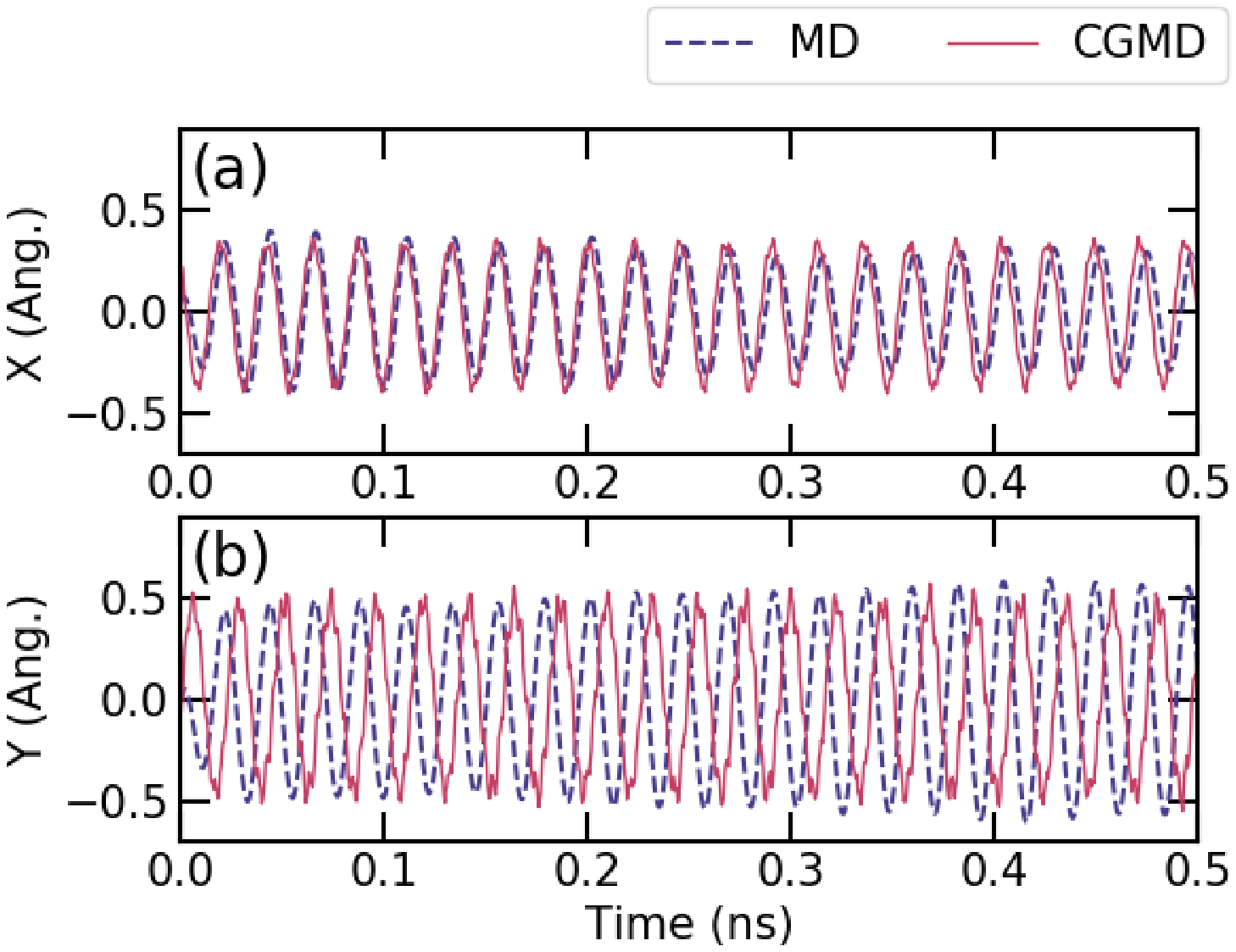}
\includegraphics[scale=0.45]{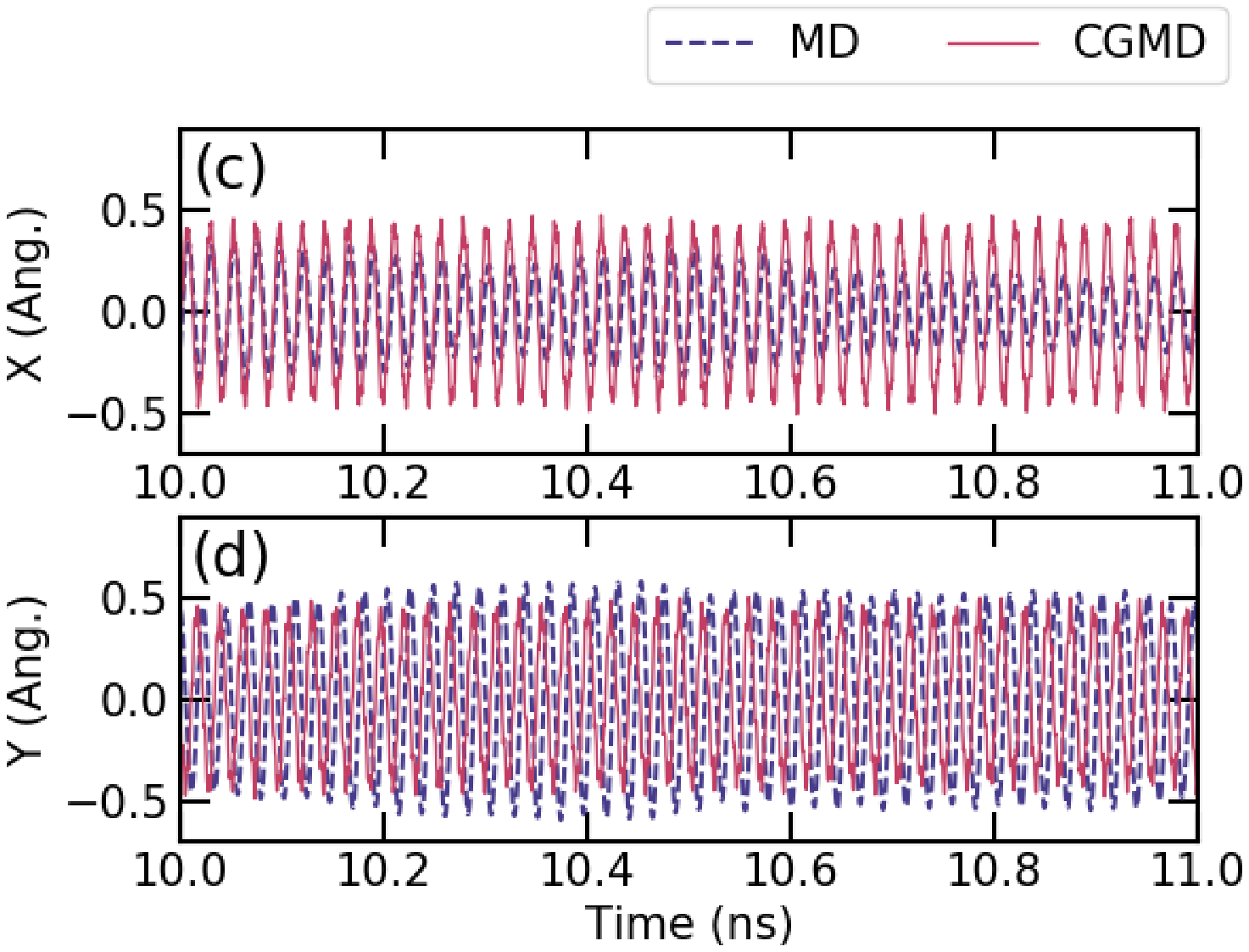}
\caption{Initial trajectory of the SWCNT calculated from MD and strain CGMD. Blue line is MD simulation and red is strain CGMD results. Strain CGMD has its initial data from MD simulation and there is no further compensation during the calculation process: (a), (b) displacements of the tip of SWNT along x and y axis for initial 0.5 ns, (c) and (d) displacements along x and y during 1 ns after 10 ns. }\label{fig:Fig5}
\end{figure}

Fig. \ref{fig:Fig5} is the trajectory of the end of the tube from CGMD compared to MD. The same time line is displayed in the animations, and animated gif is introduced in the Supporting material D. It has UA60 with soft boundary condition. For better comparison, the motion of MD simulation is included next to CGMD result in the animations. The displacements of both conditions are processed by Inverse Fast Fourier Transform (IFFT) to remove the higher frequency component. Without any additional data processing, those two simulations which share the initial condition only gives almost perfect synchronization. This displays only in a short time period, so that the result of the simulation with longer time duration is validated by counting the number of motion exchange. The detailed result is also in Supporting material D. The conclusion is that the soft boundary with 1eV fixation with LJ potential energy at the end of cantilevered beam is the most closest to the MD simulation result.

\begin{figure}
\centering
\includegraphics[scale=0.5]{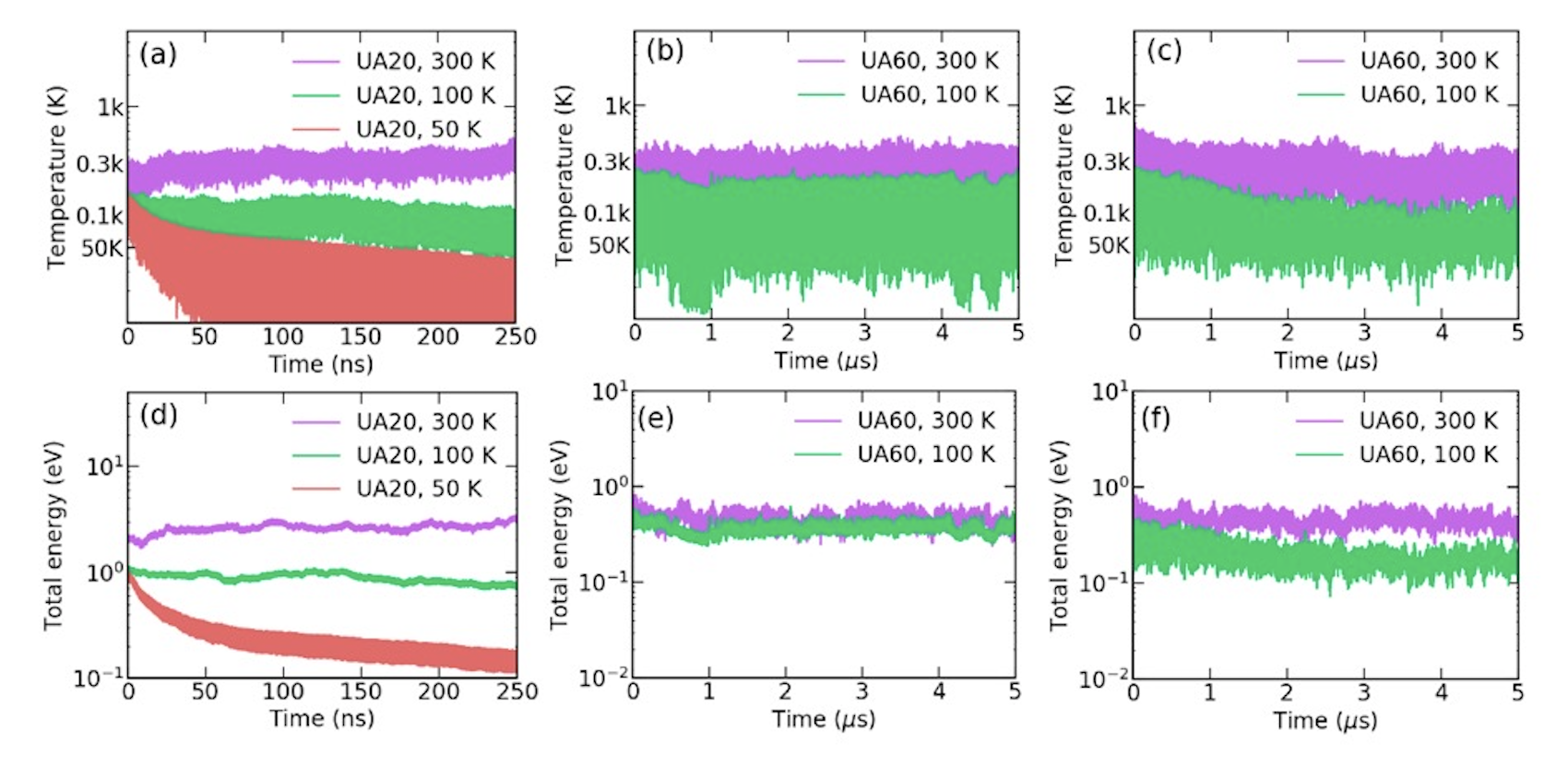} 
 \caption{Temperature calculated from spring and strain CGMD simulation without modification and Langevine thermostatting. Blue line is conventional CGMD simulation with harmonic potential function, and green is strain CGMD simulation result: (a) UA20, (b) UA60. Temperature and total energy of strain CGMD simulation without external thermostatting at 50, 100 and 300 K with red, green and magenta line, respectively: (a) and (d) UA20 with rigid boundary, (b) and (e) UA60 with 0 K rigid, boundary, (c) and (f) UA60 with LJ potential fixation with $\varepsilon$ = 1 eV for 300 K and 1 eV for 100 K.}\label{fig:Fig6}
\end{figure}

CGMD simulation calculated from the initial data of MD simulation has well resolved the thermal equilibrium condition so that each simulation with different size of beads remains with the constant temperature as shown in Fig. \ref{fig:Fig6}. Fig. \ref{fig:Fig6} (a) is UA20 case with rigid boundary. Fig. \ref{fig:Fig6} (b) is that of UA60, which also has the rigid boundary. Both results have the temperature remaining as constant level with stability, but the deviation is rather large in case of UA60. Fig \ref{fig:Fig6} (c) with LJ potential function for a fixation which is similar condition to MD simulation, shows less fluctuation.  Fig \ref{fig:Fig6} (d) $\sim$  Fig \ref{fig:Fig6} (f) are showing the total energy for each case. It is remained at constant level with given random force in Eq. (\ref{eqn:Eq13}). We do not make further argument whether the suggested calculation rigorously contents the NVE condition or ergodicity. The interest is bound to achieving a constant level of the temperature and the motion characteristics which should have the same nonlinearity as shown in MD simulation. In this reason, we would like to entitle the thermal condition that we achieved as semi-thermal equilibrium. Further study on ergodicity will be followed in another paper. The parameter set for Eq. (\ref{eqn:Eq13}) is in Supporting Information E. The simulation conditions in Fig. \ref{fig:Fig6} is introduced in Method section.

 To confirm the versatility, longer SWCNT is tested with different node length. (5,5) SWCNT of 15 nm cases is calculated through MD simulation and the results are processed as simple beads string to make the input data of CGMD simulation. The given temperature is 300 K and the same damping algorithm is applied to UA60 and UA120 with different coefficients. It gives the intended semi thermal equilibrium condition as shown in Supporting Information E. The results show the constant temperature profile with larger duration compared to other calculations. The simulations have performed under the perfect rigid fixation condition. With slightly modified parameter set, suggested algorithm shows good versatility to other simulation model with different length. 
 
The value of diffused cross correlation $\frac{ \partial ^2 p_{i,l} p_{i,\theta}}{\partial x ^2}$ during 0.5 ns along the tube axis in UA60 model with LJ potential fixation of 1 $eV$ is in Fig. \ref{fig:Fig6}. The value from MD simulation results processed as simple beads system in Fig. \ref{fig:Fig6} (a) is oscillating and that of CGMD in Fig. \ref{fig:Fig6} (b) shows similar tendency. The cross correlated state in CGMD caused by internal heat diffusion process modeled has much small range. In case of MD simulation, the amount diffused energy which is equivalent to the value of 0.1 in the Fig. \ref{fig:Fig6} (a) is around 0.15 J approximated from the applied coefficient noted in Table 1 in the Supporting Information A. This amount of energy is in oscillating during few picoseconds.  The histogram for the diffused level during whole simulation time are in Fig. \ref{fig:Fig6} (c) and (d). The frequency of appearance is strictly proportional to the amount of diffusion in log scale, which indicates the maximized diffusion between two different potential energy is rare events in free thermal vibration.  Such tendency has well mimicked in CGMD with different scale of diffusion.

For this simulation with single string, the required simulation times were approximately 2h maximum with CPU 1.6 GHz Intel Core i5.

\subsection*{Difference from the Dissipative Particle Dynamics}

Suggested algorithm in Eq.(\ref{eqn:Eq45}) $\sim$ Eq.(\ref{eqn:Eq46}) is clearly disparate from DPD modeling because it has random noise and dissipation as a single term. This algorithm is supposed to eliminate the possible evolution of artificial drift terms which is presumed to be caused by the cross correlated state between potential energy functions as being alarmed in couple of references.\cite{Peter2009,Monaghan2005} In case of MD simulation, the drift is assumed to be dissipated as small amount of perturbation in THz range so that CG modeling is managed to adapt this constrain to reduce the strong cross correlation using diffusion process. Although it is not fully elucidated in the level of phonons and second wave, almost perfect synchronized motion of CGMD and MD simulation with the constant temperature profile with the oscillating tendency of diffused energy in Fig. \ref{fig:Fig5} proves that there could be solid possibilities of more exact damping or scattering mechanism related to the diffusion.

 Modified equation of motion with cross correlation diffusion might lead the intention to several related theories which can provide better methodology to get the parameters in Eq. (\ref{eqn:Eq45}) $\sim$ Eq.(\ref{eqn:Eq46}). One powerful numerical approach would be iterative Boltzmann inversion (IBI), but this is not affordable for the parameters in Eq. (\ref{eqn:Eq45}) $\sim$ Eq.(\ref{eqn:Eq46})  because the population of the state of IBI are derived from the multiplication of partition functions which is averaging the data from the whole event separately for each potential species. Those parameters are heavily relied on various conditions such as node length, temperature and boundary rigidity. Some modification with Bayesian approach for further numerical fitting would be interesting connection. The same correlation effect has already been mentioned but it has not been quantitatively studied.\cite{Peter2009}  The parameter study based on rigorous theoretical explanation is beyond the scope of this paper. Even though the range of the parameter is conjectured from the cross correlation in Fig. \ref{fig:Fig2}, the seeking process of parameter sets are rather close to manual way.

%The autocorrelation of the velocities as given as Supporting Information D directly from each potential energy type hardly shows Markovian characteristics. In case of total velocity, specific frequency components which supposed to be from its macroscopic bending motion are included. Without this frequency range components, it is clear that the auto correlation is diminishing more rapidly than that of the velocity. 

The motion which is dealt in this paper is not in the scope of Markovian approach, yet the memory effect which includes the entropy like Zwanzig SDE \cite{Kauzlaric2011,Kauzlaric2011a} could provide more precise condition to calculate the dynamics from second sound. Thus, we strongly postulate that the suggested algorithm is working as the substitution of Langevin SDE in CG system or the friction matrix of General Langevin Equation (GLE) at this level of description. Further approach will be a theoretical proof whether the cross correlation diffusion process keeps the reversibility with constant temperature and about the formulations in free energy definition\cite{Espanol2016,Jarzynski1997,Zwanzig1961} even in more complicate molecular system. The effort to find the more fundamental mechanism with precise parameter study will give a chance to engineer the nonlinearity in nanoscale.\cite{Barnard2019,Ning2016,Tsioutsios2017} With the result of this research would help to offer more specific direction for the application on nanowire and nanotubes to be enhanced in another level.

To be more specific, the precise mechanism for the junction structure with suggested damping algorithm including the effect of ambient molecules\cite{Ma2016} like gas, liquid or collision between macromolecules are essential. For complicated molecular structure including various types of potential energy functions, there should be more number of cross correlation for each combination of the potential energy species.

\section*{Discussion}
The CG modeling of SWCNT which is able to reproduce its nonlinear dynamics has been examined. The data of displacement and velocity of each atom of (5,5) SWCNT in MD simulation is averaged into simple beads model in CGMD simulation. 8 nm and 15 nm long SWCNT are examined with three different node length at different temperature conditions in cantilevered condition. All of them have their initial condition calculated in MD simulation. We find that the comparison of the motion characteristics is not suitable with conventional CGMD algorithm because of no thermal equilibrium without random noise and the lack of complexity to reproduce nonlinear bending motion. The effect of the additional thermostat is overwhelming so that the macroscopic motion from free thermal vibration is not remaining as it is. The discrepancy between simple beads spring system and the initial condition from MD simulation is resolved using additional random force which is suggested from the diffusion of cross correlated state of bond length and angle springs kinetic energy. The achieved state of a simple bead model for SWCNT is very close to the thermal equilibrium and its macroscopic motion is proved to be compatible to nonlinear motion which is observed in atomic scale MD simulation. It also shows its advantage for longer node length with larger time step and almost perfectly synchronized motion with that from MD simulations. 
The precise reproduction of the nonlinear motion of SWCNT in CGMD simulation which is improved has shown that the causality of nonlinear motion of SWCNT  is the internal heat diffusion in THz range.

\section*{Methods}
\subsection*{MD and CGMD simulation}
The appearance of nonlinear motion of SWCNT in thermal equilibrium depends on temperature and aspect ratio. Long SWCNT has less motion exchange. If the motion exchange is too rare, the examination of suggested model will be inefficient. Short SWCNT makes the motion too noisy with extreme complexity so that the validation will not be easy. Choosing proper SWCNT and temperature before the simulation ensures the convenient numerical experiments. Under these considerations, (5,5) SWCNT with 8 nm long at 300 K is modeled into simple beads system. In the former MD simulation study,\cite{Koh2015} the trend of motion exchange clearly depends on the rigidity of the fixed end boundary condition. The fixed end with Lennard-Jones (LJ) potential function is employed to make sure of the affordability of suggested CGMD simulation. This SWCNT at 300 K provides exemplary nonlinear characteristic dynamics as Fig. \ref{fig:Fig1}. 

 The MD simulation was performed with the LAMMPS package\cite{Plimpton1995} with adaptive intermolecular reactive bond order (AIREBO) potential function\cite{Stuart2000} with time step of 0.5 $fs$. Langevin thermostat with damping coefficient 0.01 ps is attached during the initial 1 ns. The displacement and velocity data of all atoms are captured after 1 $ns$ of relaxation. The rigid end fixation is applied at the bottom with phantom wall condition which has the length scale $\sigma$ of 0.89 \AA with LJ potential function. The rigidity of the end fixation, which is decided with the LJ energy scale is applied for CGMD simulation as well. For comparison, the fixation rigidity has given with 1 and 5 eV. 
 
  To define CGMD beads, the displacement and velocity of every 60 atoms are simply averaged as one lumped mass, i.e. a bead. The target SWCNT has 660 atoms so that the system has 11 of lumped mass. It is regarded as a unit atom (UA), which is equivalent to a coarse grained (CG) particle. To make proper bending motion, however, one additional unit atom should be attached next to the fixed end of the simple beads model in CGMD simulation. In this way, the bending angle between fixed end and neighbours can be managed. Additionally, simple beads systems with lumped mass consisted of 20 and 120 atoms are also tested. The bond length for CG particle incorporating 20 carbon atoms (UA20) is 2.42  \AA and the mass is 240 amu. Those of CG particle for 60 and 120 atoms (UA60 and UA120, respectively) is calculated as well as the case of UA20.
  
  The force constants for bond length and angle spring are $k_{sp}$ = 220 eV/Å  and $k_{ang}$ = 2200 eV \AA $\sim$ 2800 eV \AA  from the parameter study\cite{Zhigilei2005a} of each spring which comes from the rigorous measurement using external force at 0 K. The precise value of $k_{ang}$ is decided by the peak location in the frequency domain. Velocity Verlet algorithm has been employed with the time step of 0.5 $fs$ for the case of UA20. Time step of 10 $fs$ is adapted for UA60 and UA120. Total time spans are 250 $ns$ and 5 $\mu$s for UA20 and UA60/UA120, respectively.

\subsection*{Cross correlation condition}
  In Fig. \ref{fig:Fig6}, the difference of the strain, $\Delta_t L$ and angle, $\Delta_t \Theta$, in 50 fs of an arbitral node of a string, which is equivalent to $p_i^\ell$and $p_i^\theta$ with $1/\Delta t$ respectively, are sampled from MD and CGMD simulation and processed to show cross correlation in frequency domain. In case of MD simulation, the value of strain and angle of beads model is averaged from atomic structure of SWCNT for each node. The cross correlation during 500 ps is used for FFT. Most of the peaks appeared in THz range for both of CGMD and MD simulation. No significant results have been found in the range below THz. The CGMD simulation has more distinctive peaks than that of MD simulation results. This means 1) CGMD has the velocity values caused by angle and bond length which are severely correlated in periodic manner because there is no constrain and 2) MD simulation has cross correlation which is not active in the bending frequency range but in optical mode range. In MD simulation, the force caused by $2\bm{\ell_i} \bm{\theta_i}$ is close to small perturbation in CG description level, and it could be disappeared as internal heat diffusion process as fluctuation-dissipation in THz range.

 \subsection*{Acknowledgements}
 Part of this work was supported by JSPS KAKENHI Grant Numbers JP15H05760, JP18H05329 and by the Brain Korea 21 Plus Project in BK21 Plus Transformative Training Program for Creative Mechanical and Aerospace Engineers. We sincerely appreciate to Prof. Hyeon at KIAS for his valuable comments and discussion. The code for UA60 at 300 K with soft fixation is in \href{https://github.com/ieebon/Strain_CGMD}{Github} or go to the next url: \url{http://github.com/ieebon/Strain_CGMD}.

%\showmethods{}
%\showacknowledgements{}
      % APS-like style for physics
\bibliography{main.bib}
\end{document}